# Spectrum multiplexing and coherent-state decomposition in Fourier ptychographic imaging

Siyuan Dong,[1] Radhika Shiradkar,[1] Pariksheet Nanda,[1] and Guoan Zheng[1,2*]

[1]Biomedical Engineering, [2]Electrical and Computer Engineering, University of Connecticut, Storrs, CT, 06269, USA
*guoan.zheng@uconn.edu

**Abstract:** Information multiplexing is important for biomedical imaging and chemical sensing. In this paper, we report a microscopy imaging technique, termed state-multiplexed Fourier ptychography (FP), for information multiplexing and coherent-state decomposition. Similar to a typical Fourier ptychographic setting, we use an array of light sources to illuminate the sample from different incident angles and acquire corresponding low-resolution images using a monochromatic camera. In the reported technique, however, multiple light sources are lit up simultaneously for information multiplexing, and the acquired images thus represent incoherent summations of the sample transmission profiles corresponding to different coherent states. We show that, by using the state-multiplexed FP recovery routine, we can decompose the incoherent mixture of the FP acquisitions to recover a high-resolution sample image. We also show that, color-multiplexed imaging can be performed by simultaneously turning on R/G/B LEDs for data acquisition. The reported technique may provide a solution for handling the partially coherent effect of light sources used in Fourier ptychographic imaging platforms. It can also be used to replace spectral filter, gratings or other optical components for spectral multiplexing and demultiplexing. With the availability of cost-effective broadband LEDs, the reported technique may open up exciting opportunities for computational multispectral imaging.

## 1. Introduction

Fourier ptychography (FP) [1] is a recently developed microscopy imaging approach that applies angular diversity functions for high-resolution complex image recovery. Sharing its root with synthetic aperture technique [2, 3], FP synthesizes many variably illuminated, low-resolution intensity images in the Fourier space to expand the passband and surpass the diffraction limit of the objective lens. The recovery process of FP switches between spatial and Fourier domains. In spatial domain, the amplitudes of the acquired images are used to constraint the FP solution, similar to the strategy of phase retrieval techniques [4-14]. In the Fourier domain, the panning Fourier constraints are imposed to reflect the angle-varied illuminations. The name 'Fourier ptychography' comes from ptychography [15-27], a lensless imaging technique originally proposed for transmission electron microscopy [26]. Both FP and lensless ptychography adopt phase diversity concept [8, 11] for image recovery. Lensless ptychography applies translational diversity (laterally moving the sample) [16] for inverting the diffraction process and recovering the complex image without using lens elements. Fourier ptychography, on the other hand, introduces the 'angular diversity' concept to simultaneously expand the Fourier passband and recover 'lost' phase information [28]. The use of lens elements in FP settings also provides a higher signal-to-noise ratio and reduces the coherence requirement of the light source. It has been shown that, without involving any interferometry measurement or mechanical scanning, FP surpasses the diffraction limit defined by the numerical aperture (NA) of the objective lens [1].

Despite the successful demonstration of the FP approach, its operation is currently limited to the single coherent state of the light source. In other words, the light source in FP settings is assumed to be spatially a point source and of temporally a single wavelength. Incoherent mixture of multiple coherent states has not been considered in the FP recovery procedures. Therefore, one important step for advancing the FP technique is to develop a recovery scheme for handling state mixture and performing information multiplexing with FP acquisitions. As we will discuss later, such a state-multiplexed FP scheme may find applications in coherent-state decomposition and computational multispectral imaging.

Recently, mode expansion of the mutual coherence function has been reported for coherent diffractive imaging, which allowed reconstruction using partially coherent light sources [29-31]. A mixed-state formulation has also been reported for the lensless ptychography approach by Thibault and Menzel [32], and was recently applied to information multiplexing by Batey *et.al* [33]. Motivated by the previous works, in particular, the mixed-state formulation of lensless ptychography [32], we report a state-multiplexed recovery scheme for Fourier ptychographic imaging settings. We validate the reported scheme with both simulations and experiments.

This paper is structured as follows: in Section 2, we describe the recovery procedures of the state-multiplexed FP scheme. In Section 3, we demonstrate the effectiveness of the reported scheme with both simulations and experiments. In Section 4, we demonstrate the use of the reported scheme for color-multiplexed FP imaging. Finally, we summarize the results and discuss future directions.

## 2. State-multiplexed Fourier ptychography

To understand the operation of the state-multiplexed FP scheme, it is worthwhile to review the basic concepts of single-state FP approach. As detailed in Ref. [1], a typical FP platform consists of an LED array, a conventional microscope with a low-NA objective lens, and a monochromatic CCD camera. The LED elements on the array are turned on sequentially to illuminate the sample from different incident angles. At each illumination angle, the camera acquires a low-resolution intensity image of the sample. These acquired images are then stitched with overlap in the Fourier domain using the single-state FP algorithm. The single-state algorithm starts with a high-resolution estimate of the sample profile (can be a random guess): $I_h e^{i\varphi_h}$. This sample estimate is then low-pass filtered to produce a low-resolution target image $I_t e^{i\varphi_t}$, simulating the filtering process of the objective lens. Next, the intensity component of the target image $I_t$ is replaced by the actual measurement $I_m$, while the phase component is kept unchanged. The updated target image is then used to modify the corresponding spectrum region of the sample estimate. This replace-and-update sequence is repeated for all intensity measurements, and iterated several times until the solution converges.

In such a single-state FP algorithm, the generated low-resolution target images uniquely map to different regions of the sample estimate in the Fourier space. This one-to-one mapping relationship is a direct consequence of the single-coherent-state assumption, i.e., the illumination is assumed to be a point source spatially and that of a single wavelength temporally. In the case of low-coherent Fourier ptychographic acquisition, the intensity measurement represents an incoherent summation of different coherent states [32], and thus, the mapping between the low-resolution target image and the high-resolution sample estimate is not in a one-to-one relationship. In the reported state-multiplexed FP scheme, we model the decoherent effect using multiple target images, corresponding to different coherent states of the light source [32].

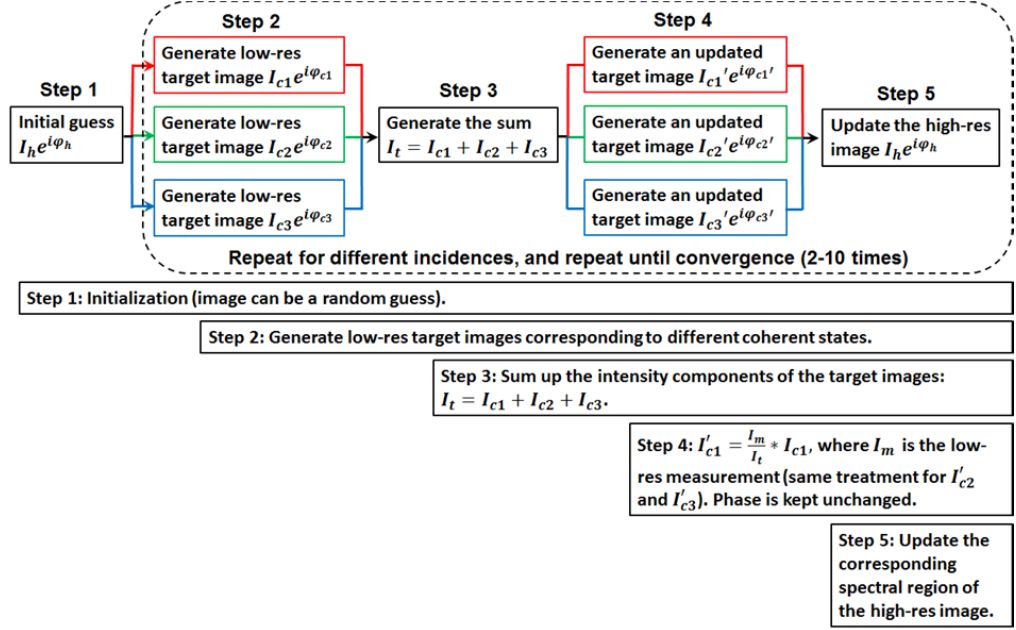

Fig. 1. The recovery procedures of the state-multiplexed Fourier ptychography scheme.

The recovery process of the state-multiplexed FP scheme is shown in Fig. 1. Similar to the single-state scheme, it starts with a high-resolution estimate of the sample profile: $I_h e^{i\varphi_h}$. This sample estimate is used to generate multiple low-resolution target images corresponding to different coherent states. Second, the intensity components of the target images are summed up to generate the incoherent mixture $I_t$. Third, the ratio between the actual measurement $I_m$ and $I_t$ is used to update the intensity components of the target images, while the phase components are kept unchanged. Fourth, the updated target images are used to modify the corresponding spectral regions of the sample estimate. Lastly, the entire process is repeated for all intensity measurements, and iterated for several times until the solution converges.

The key difference between the reported scheme and the single-state FP lies in the intensity replacement process. In the single-state FP, the intensity component of the target image is directly replaced by the actual measurement $I_m$ while the phase component is kept unchanged. The reported state-multiplexed scheme, on the other hand, uses the ratio between the incoherent mixture and the actual measurement to update the intensity components of the target images. This new updating process ensures that the intensity summation of different coherent modes equates to the measured incoherent mixture, while the phase of individual modes is preserved.

## 3. Simulations and experiments of the state-multiplexed FP scheme

We first validated the state-multiplexed scheme using simulations. The simulation parameters were chosen to realistically model a light microscope experiment, with an incident wavelength of 632 nm, a pixel size of 1.375 μm (at the sample plane) and an objective NA of 0.1 (Olympus 4X Plan). We simulated the use of a 15*15 LED (SMD 3528) array for illuminating the sample with different incident angles. The LED array is placed 85 mm beneath the sample, and the distance between adjacent LEDs is 4 mm.

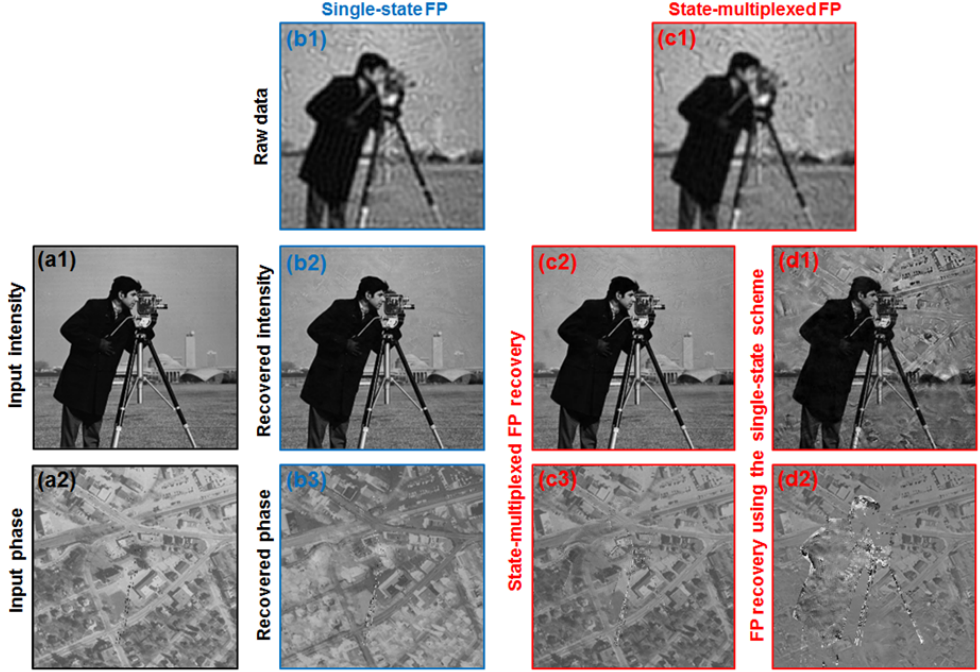

Fig. 2 Simulations of the single-state and state-multiplexed FP schemes. (a1) and (a2) The input intensity and phase images of the simulated object. (b1) Raw data of the single-state FP scheme. Each low-resolution image (0.1 NA) corresponds to one LED element at the array. (b2) and (b3) The recovered sample intensity and phase images using the single-state FP scheme. (c1) Raw data of the state-multiplexed FP scheme. Two adjacent LED elements are lit up simultaneously for sample illumination, and each low-resolution intensity image (0.1 NA) represents an incoherent summation of two coherent states. (c2) and (c3) The recovered sample intensity and phase images (0.5 NA) using the state-multiplexed FP scheme. (d1) and (d2) The reconstructions of state-mixed raw data using the single-state FP scheme (for comparison).

The high-resolution input intensity and phase profiles are shown in Fig. 2(a1) and (a2), which serve as the ground truth for the simulated complex object. We then simulated the low-resolution measurements for 1) the single-state FP scheme and 2) the state-multiplexed FP scheme. For the single-state scheme, each low-resolution intensity image corresponds to one LED element in the array (225 images), as shown in Fig. 2(b1). The corresponding reconstructed intensity and phase images are shown in Fig. 2(b2) and 2(b3). For the state-multiplexed scheme, we simulated the case of two adjacent LED elements lighting up simultaneously for sample illumination. In this case, each low-resolution intensity image (Fig. 2(c1)) represents an incoherent summation of two coherent states. As we group two LEDs as one light source element, the total number of simulated low-resolution images reduces by half (113 images; we did not group the last LED element). We note that, the spectrum-overlapping percentage (i.e., the data redundancy requirement) is determined by the distance between two adjacent light sources. In this case, two LEDs can be viewed as one 'larger' light source element, and the resulting averaged spectrum-overlapping percentage is determined to be

~55%, satisfying the convergence condition of ~40% [34]. Based on the simulated low-resolution state-mixtures, we applied the state-multiplexed FP scheme to recover the high-resolution sample images in Fig. 2(c2) and 2(c3). Using the state-mixture data set, we also used the single-state FP scheme (for comparison purpose) to reconstruct the sample images in Fig. 2(d1) and 2(d2).

The image qualities of different FP reconstructions are quantified using the mean-square errors (the difference between the FP reconstructions and the ground truth; smaller errors represent better reconstruction). The mean-square errors for Fig. 2(b2), 2(c2), and 2(d1) are 0.11%, 0.47% and 4.0%, respectively. From this set of simulations, we can see that, the reported state-multiplexed scheme is able to recover the sample image with quality similar to single-state FP.

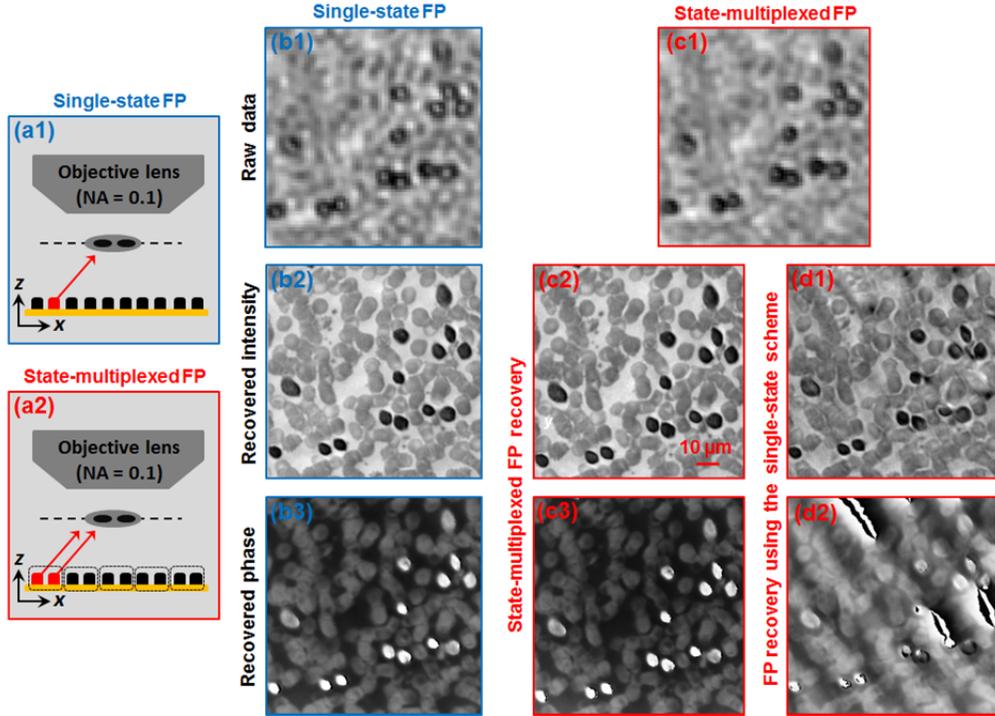

Fig. 3 Experiments of the single-state and state-multiplexed FP schemes. (a1) and (a2) The experimental setups for the two schemes. (b1) Raw data of the single-state FP scheme (0.1 NA). (b2) and (b3) The recovered sample intensity and phase images using the single-state FP scheme (0.5 NA). (c1) Raw data of the state-multiplexed FP scheme (0.1 NA). Two adjacent LED elements are lit up simultaneously for sample illumination, and each low-resolution intensity image represents an incoherent summation of two coherent states. (c2) and (c3) The recovered sample intensity and phase images using the state-multiplexed FP scheme (0.5 NA). (d1) and (d2) The reconstructions of state-mixed raw data using the single-state FP scheme (for comparison).

We next validated the state-multiplexed FP scheme using a light microscope experiment, as shown in Fig.3. The parameters of the experimental setup are similar to those of simulations. For the single-state FP scheme (Fig. 3(a1)), each acquired image corresponds to illumination from one LED element. Fig. 3(b1) shows an example of the single-state low-resolution raw image. For the state-multiplexed FP scheme (Fig. 3(a2)), two adjacent LEDs are used for single image acquisition, and thus, the corresponding raw image (Fig. 3(c1)) represents an incoherent summation of two coherent states (from two LEDs). The reconstructed high-resolution images for these two schemes are shown in Fig. 3(b2-b3) and 3(c2-c3). Based on the state-mixed raw images, we also used the single-state FP scheme to reconstruct the high-resolution images in Fig. 3(d1-d2) for comparison.

From Fig. 3, we can see that, the state-multiplexed FP is able to recover high-resolution sample images from state-mixed measurements, and the quality of the reconstructions is comparable to that of single-state FP. This experiment validates the effectiveness of the reported state-multiplexed FP recover routine.

## 4. Spectrum multiplexing in Fourier ptychographic imaging

One important application of the state-multiplexed FP scheme is multispectral imaging. Light sources with multiple wavelengths can be used to illuminate the sample from different incident angles, and the acquired images represent incoherent summations of the sample transmission profiles at different wavelengths. A state-multiplexed FP algorithm can then be used to reconstruct multiple high-resolution images at different wavelengths. In this section, we demonstrate such a state-multiplexed scheme for color multiplexing, which we will refer to as "color-multiplexed FP". It is also important to acknowledge that, color multiplexing has recently been demonstrated for lensless ptychography by using the concept of translational diversity (laterally moving the sample over different spatial positions) [33].

The operation principle of color-multiplexed FP is shown in Fig. 4, where we used a color LED array for sample illumination and acquired low-resolution images using a monochromatic camera. Since R/G/B LED elements were turned on simultaneously in the acquisition process, the acquired images represent incoherent summations of sample profiles at R/G/B wavelengths. Such state-mixed raw data is then used to recover three high-resolution images at the corresponding wavelengths and produce a final color image of the sample.

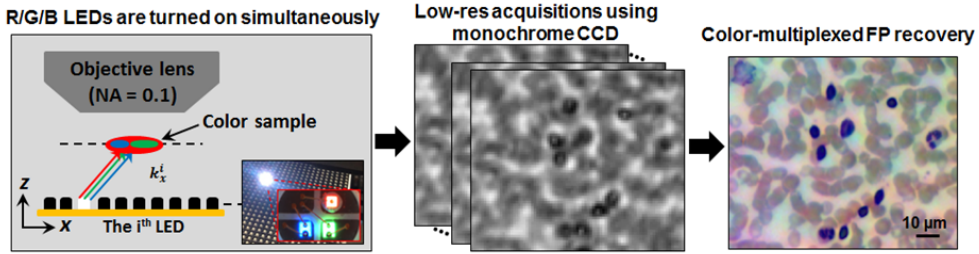

Fig. 4 Color-multiplexed FP scheme. R/G/B LEDs are turned on simultaneously for illumination. Low-resolution images are acquired using a 0.1 NA objective lens and a monochrome camera. A color-multiplexed FP recovery algorithm is then used to decouple the R/G/B channels from the low-resolution images. A high-resolution color image of the sample can be recovered using computation instead of spectral filters.

The recovery process of the color-multiplexed FP scheme is similar to that of state-multiplexed scheme. However, in this case, multiple sample estimates at different wavelengths are used in the workflow. As shown in Fig. 5, three sample estimates are used to generate the corresponding target images. The intensity components of the target images are summed up to generate the incoherent mixture $I_t$, and the target images are updated using the ratio between the actual measurement $I_m$ and $I_t$. The updated target images are then used to modify the corresponding spectral regions of the sample estimates. The entire process is repeated for all intensity measurements, and iterated for several times until the solution converges. Lastly, the recovered images at different wavelengths are combined to produce a high-resolution color image.

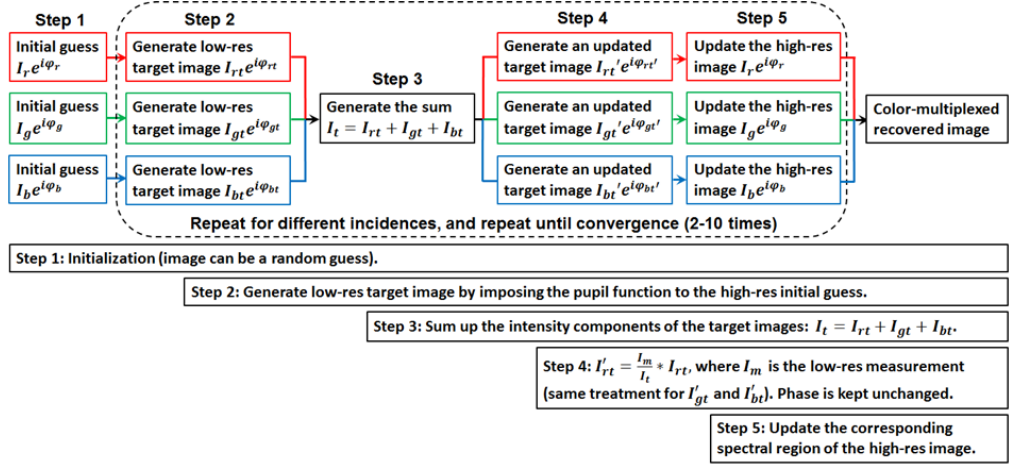

Fig. 5 Color-multiplexed FP recovery routine.

As before, we first validated the color-multiplexed FP using simulations. We simulated a 15*15 color LED array with red/green/blue (632 nm, 532 nm, and 472 nm) elements lighting up simultaneously; other parameters being the same. The high-resolution input images (the ground truth) at three different channels are shown in Fig. 6(a1-a3), and the corresponding color image is shown in Fig. 6(a4). The simulated low-resolution measurement is shown in Fig. 6(b), which represents an incoherent summation of three coherent states of the sample profiles. We then applied the color-multiplexed FP recovery algorithm to perform reconstruction, and the results are shown in Fig. 6(c1-c4). We also quantified the image quality by calculating the mean-square errors, and the results are 0.5%, 0.4%, and 0.1% for Fig. 6(c1-c3) respectively. We can see that, the color-multiplexed FP scheme is able to recover the color image of the sample from state-mixed measurements.

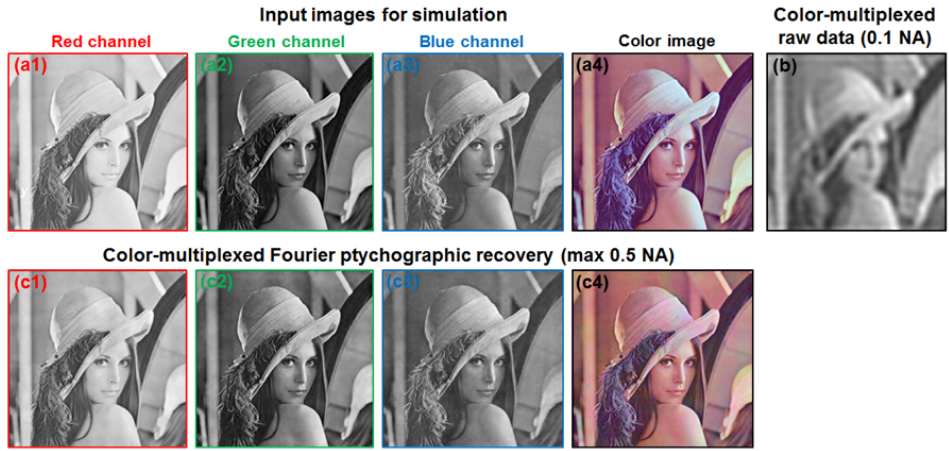

Fig. 6 Simulations of the color-multiplexed FP scheme. (a) Input R/G/B channels and the color image. (b) The low-resolution intensity measurement of the object, representing an incoherent summation of 3 object profiles at different wavelengths. (c) The color-multiplexed FP recovery (0.5 NA). The mean-square errors for (c1)-(c3) are 0.5%, 0.4%, and 0.1%, respectively.

We also validated the color-multiplexed FP scheme using a light microscope experiment, as shown in Fig. 7. We used a pathology slide (human breast cancer section, Carolina Inc.) as our sample, and R/G/B LEDs were turned on simultaneously for sample illumination. Low-resolution images were acquired using the 0.1 NA objective lens and the monochrome camera (parameters are the same as that for simulations). As such, these measurements represent incoherent mixtures of sample profiles at three wavelengths, as shown in Fig. 7(a). The color-multiplexed FP algorithm was then used to decouple into the R/G/B channels from the state-mixed measurements and recover the high-resolution color images of the sample. Fig. 7(b1)-7(b3) demonstrate the recovered images for R/G/B channels, and the final color image is shown in Fig. 7(c). We also reconstruct the color image from three separated FP acquisitions (without state-mixing) in Fig. 7(d). As a comparison, the color image captured using a conventional 40X objective lens (0.6 NA) is shown in Fig. 7(e). However, we note that, the focal plane of Fig. 7(e) may slightly shift when we switch the objective lens (the sample is about 15 μm thick).

From this experiment, we can see that, the color-multiplexed FP algorithm is able to recover the high-resolution color image from the state-mixed measurements. We found that, the major noise source in our experiment was from the intensity uncertainty of the light source array. For example, the intensity fluctuations of R/G/B LED elements change the ratio of color components, creating state-decomposing errors and color mismatch problem for FP reconstructions. To address this problem, we can use a spectroscopic element to better calibrate the intensity values of the different LEDs in the array. We also found that, under the same illumination intensity, the intrinsic image contrast for the blue channel is weaker that of red and green channels. To address this problem, we can adjust the illumination intensity ratio between different wavelength components. The optimal intensity ratio may depend on the employed staining technique. Nevertheless, the experiment shown in Fig. 7 provides a proof-of-concept demonstration of the color-multiplexed FP scheme.

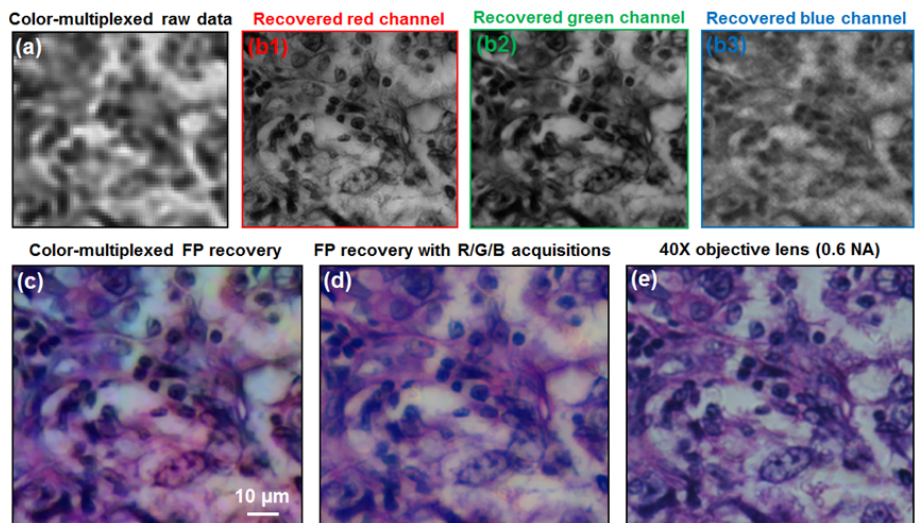

Fig. 7 Experimental demonstration of the color-multiplexed FP scheme. (a) Raw data of the color-multiplexed FP acquisition, representing incoherent summation of the sample profiles at three wavelengths. (b1)-(b3) The recovered color-multiplexed high-resolution images (0.5 NA) at red, green, and blue channels. (c) The recovered color image by combining (b1)-(b3). (d) The recovered color image (0.5 NA) using three separated FP acquisitions with individual red, green, and blue illumination (no state-mixing). (e) Color image using a conventional microscope with a 40X high-NA objective lens (0.6 NA).

## 5. Conclusion

In conclusion, we have developed and tested a state-multiplexed recovery scheme for Fourier ptychographic imaging. In this scheme, we use multiple target images for handling different coherent states of the light source and then update the high-resolution sample image accordingly. We have demonstrated the applications of the reported scheme for coherent-state decomposition and color-multiplexed imaging.

The state-multiplexed scheme can be used to model the partially coherent effects of the employed light sources. For example, the finite extent of the light source (related to spatial coherence) can be modeled as multiple point sources emitting light independently. The finite spectrum of the light source (related to temporal coherence) can be modeled as multiple light sources emitting light with different, but narrower passbands. However, we note that, in our current implementation, the major reconstruction error comes from the intensity uncertainty of the LED array, not the partially coherent effects of the light source element. We also note that, instead of light up two components simultaneously, we can also light up more components in the array to form, for example, Hadamard basis patterns. By using the optimal basis, we may be able to reduce the number of acquisitions and raise the photon budget.

On the other hand, the color-multiplexed scheme reported in this work can be used to replace thin-film interference filters, gratings or other optical components for spectral multiplexing and demultiplexing. With the availability of cost-effective broadband LEDs, the reported scheme may open up exciting opportunities for computational multispectral or even hyperspectral imaging. Our on-going efforts include the development of a high-power white light LED array for sample illumination. Finally, we note that, the effectiveness of the spectrum-multiplexed approach may depend on the data redundancy and the compressibility of the sample image [34]. This relationship can be directed to the recent development of compressive sensing [35-38]. Further investigations along this line are highly desired.

## Acknowledgments

We are grateful for the discussions with Mr. Xiaoze Ou and Dr. Ying Min Wang. For more information, please visit https://sites.google.com/site/gazheng/.